# STOCHASTICALLY BUNDLED DISSIPATORS FOR THE QUANTUM MASTER EQUATION

SAYAK ADHIKARI AND ROI BAER*

*Fritz Haber Center for Molecular Dynamics and Institute of Chemistry, The Hebrew University of Jerusalem, Jerusalem 9190401, Israel*

Abstract. The Lindblad master equation is a fundamental tool for describing the evolution of open quantum systems, but its computational complexity poses a significant challenge, especially for large systems. This article introduces a stochastic representation of the Lindblad dissipator that addresses this challenge by bundling the Lindblad operators. We demonstrate the effectiveness of this method by considering a Morse oscillator coupled to a spin bath. Our numerical experiments show that a small number of stochastically bundled operators can accurately capture the system's dynamics, even when the Hilbert space dimension is large. This method offers a new perspective on open quantum systems and provides a computationally efficient way to simulate their dynamics.

## 1. Introduction

Quantum systems, by their nature, interact with their environment, making them inherently "open" to external influences. In many applications, their dynamics can be effectively modeled using Markovian master equations applied to the system's density matrix [1–4]. The Lindblad Master Equation (LME) [5–10] is the fundamental Markovian quantum master equation used to describe this evolution. Independently developed by Lindblad [11] and Gorini, Kossakowski, and Sudarshan [12], the LME is given by:

$$\dot{\rho}\left(t\right)=\frac{1}{i\hbar}\left[\mathcal{H},\rho\left(t\right)\right]+\mathscr{D}\rho\left(t\right), \tag{1.1}$$

where $\mathcal{H}$ is the effective Hermitian Hamiltonian, and the quantum dissipator $\mathscr{D}$ has the form:

$$\mathscr{D}\rho\equiv\frac{1}{2}\sum_{\omega\in B}\gamma\left(\omega\right)\left(\left[\mathcal{L}_{\omega}\rho,\mathcal{L}_{\omega}^{\dagger}\right]+\left[\mathcal{L}_{\omega},\rho\mathcal{L}_{\omega}^{\dagger}\right]\right). \tag{1.2}$$

Here, $B$ is a set of indices $\omega$, $\gamma\left(\omega\right)$ are non-negative coupling functions and the Lindblad operators $\mathcal{L}_{\omega}$ capture the effects of the environment on the system. The LME preserves the positivity and unit trace of the reduced density operator, crucial for probability calculations.

In chemical physics researchers often use the LME to study energy transfer, relaxation processes, and chemical reactions [13–20]. It is also utilized in condensed matter physics to investigate transport phenomena, non-equilibrium dynamics, and the behavior of disordered and strongly correlated systems [21–25]. Within quantum information science, it plays a crucial role in modeling decoherence in qubits, analyzing quantum gates, and characterizing noisy communication channels [26–28]. Furthermore, it is essential for advancements in quantum thermodynamics, where it elucidates fundamental limits on energy conversion and information processing [9, 29–31].

Beyond these specific applications, the LME has also fundamentally changed our understanding of open quantum systems. It expresses the idea that the evolution of a pure state is shaped by both the system's own internal dynamics and its random interactions with the environment [32, 33]. In fact, the Lindblad operators themselves have been interpreted as representing random measurements performed by the environment on the system [6, 34–40].

Stochastic approaches like quantum jump [41–44] and quantum state diffusion [45, 46] provide a way to sample these random system trajectories over time, and when averaged, these trajectories yield the Lindbladian density matrix.

Davies [47, 48] derived the form of the effective Hamiltonian and Lindblad operators for systems weakly coupled to a thermal environment at temperature $T$, described by the density matrix $\rho_E(T)$. In this formalism, the system Hamiltonian is $\mathcal{H}_{sys}$, and the system-environment coupling is $H_{s-E} = \mathcal{X} \otimes \mathcal{Y}$, where $\mathcal{X}$ and $\mathcal{Y}$ are system and environment operators, respectively. The set $B$ corresponds to the distinct Bohr frequencies $\omega$ arising from energy level differences in $\mathcal{H}_{sys}$. The effective Hamiltonian is:

$$\mathcal{H} = \mathcal{H}_{sys} + \mathcal{H}_{Ls}, \tag{1.3}$$

with the Lamb-shift operator:

$$\mathcal{H}_{Ls} = \sum_{\omega \in B} \Im \Gamma_{\mathcal{Y}\mathcal{Y}}(\omega)\, \mathcal{L}_\omega^\dagger \mathcal{L}_\omega, \tag{1.4}$$

where

$$\Gamma_{\mathcal{Y}\mathcal{Y}}(\omega) = \int_0^\infty e^{i\omega t} \mathrm{Tr}_E[\rho_E(T)\, \mathcal{Y}(t)\, \mathcal{Y}] dt \tag{1.5}$$

is the Fourier transform of the environment operator's autocorrelation function, and the Lindblad operators are:

$$\mathcal{L}_\omega = \sum_{nm} |n\rangle\langle n|\mathcal{X}|m\rangle\langle m| \delta_{\hbar\omega, \varepsilon_m - \varepsilon_n}, \tag{1.6}$$

where $\varepsilon_n$ and $|n\rangle$ are, respectively, the eigenvalues and eigenkets of $\mathcal{H}_{sys}$. Davies also gave the explicit form for the coupling function $\gamma(\omega) = 2\Re\Gamma_{\mathcal{Y}\mathcal{Y}}(\omega) \geq 0$, noting it satisfies the detailed balance condition: $\gamma(\omega) = e^{\hbar\omega/k_BT}\gamma(-\omega)$, ensuring that the LME drives the system to thermal equilibrium:

$$\lim_{t\to\infty} \rho(t) = \rho_{eq}(T) \equiv \frac{e^{-\mathcal{H}/k_BT}}{\mathrm{Tr}[e^{-\mathcal{H}/k_BT}]}, \tag{1.7}$$

as required for systems weakly coupled to the environment.

While the Davies formalism provides a valuable framework for deriving the LME, its practical application to large quantum systems faces significant computational hurdles. Before tackling the computational challenges of applying Lindblad equations to large systems, their validity in this regime must be established. The Davies approach, often used to derive Lindblad operators (Eq. 1.6 of the main text), is limited when applied to systems with a high density of states—such as nanocrystals, semiconductors, and light-harvesting complexes—because it relies on the secular approximation, which assumes well-separated energy levels. However, recent developments have yielded alternative methods that transform the Redfield equation into a Lindblad form without requiring the full secular approximation [49–54]. These methods validate the use of the LME for studying open quantum systems with dense state spaces.

Given the validity of the LME for these systems, we now turn to the computational challenges associated with its numerical solution. These challenges have spurred the development of specialized software packages, libraries, and stochastic algorithms [55–58]. The computational complexity scales dramatically with the Hilbert space dimension. This scaling arises from the increasing number of Lindblad operators needed to describe the system's interaction with its environment. Each operator represents a potential pathway for the system to exchange energy or information with the environment, leading to a complex dissipator term in the LME.

To illustrate these challenges, consider a Hilbert space of finite dimension $N$. Evolving the $N \times N$ density matrix $\rho$ via the LME involves operations with matrices of the same dimension representing the Hamiltonian and Lindblad operators. Assuming no special structure or sparsity, and recognizing that the number of Lindblad operators scales as $\mathcal{O}(N^2)$ [6], we analyze the complexity of two common approaches. First, consider direct application of the dissipator. Preparing the Lindblad matrices requires $\mathcal{O}(N^4)$ operations. Applying the dissipator to $\rho$ involves $\mathcal{O}(N^2)$ matrix multiplications (one for each Lindblad operator), each with a complexity of $\mathcal{O}(N^3)$, resulting in an overall scaling of $\mathcal{O}(N^5)$ per time step. A second approach utilizes the Lindbladian superoperator, an $N^2 \times N^2$ matrix. Constructing this superoperator is formally $\mathcal{O}(N^6)$ but is performed only once. Subsequent propagation by applying the superoperator to $\rho$ requires $\mathcal{O}(N^4)$ operations per time step.

This high computational cost underscores the need for efficient numerical methods. These challenges can arise even for systems with a relatively small number of degrees of freedom if they are strongly coupled to their environment. In such scenarios, a common strategy is to incorporate relevant environmental modes (e.g., reaction coordinates) into the system Hamiltonian. By renormalizing environmental correlation functions, this approach effectively transforms a strongly coupled small system into a weakly coupled larger system, often leading to more tractable calculations [59, 60]. This transformation further motivates the development of optimized numerical techniques for large systems.

This paper introduces a novel method for addressing the computational bottleneck arising from the quadratic growth in the number of Lindblad operators with increasing system size. We demonstrate that the standard Lindblad dissipator can be transformed into a stochastic dissipator containing a fixed number of "bundled" Lindblad operators, significantly reducing the complexity of the calculations while preserving the essential Lindblad form and maintaining accuracy within a specified error threshold for relevant observables.

## 2. Method

2.1. **Stochastic bundled dissipators.** Our procedure involves a vector $\boldsymbol{r} = (r^{\omega})_{\omega \in B}$, whose components are real or complex independent random variables, each with zero expected value and unit variance:

$$\mathbb{E}r^{\omega}r^{\omega'*} = \delta_{\omega\omega'} \to \mathbb{E}\boldsymbol{r}\boldsymbol{r}^{\dagger} = I. \tag{2.1}$$

For example, $r^{\omega}$ can be discrete random variables sampling from $\{-1, 1\}$ or a random number on the unit circle of the complex plane. We then form the stochastically bundled Lindblad operator

$$\mathcal{R}_1 = \sum_{\omega \in B} r^{\omega} \sqrt{\gamma(\omega)} \mathcal{L}_{\omega}.$$

with which we can define the random dissipator $\mathscr{D}_1$ as a formal operation:

$$\mathscr{D}_1 \rho \equiv \frac{1}{2}\left(\left[\mathcal{R}_1 \rho, \mathcal{R}_1^{\dagger}\right] + \left[\mathcal{R}_1, \rho \mathcal{R}_1^{\dagger}\right]\right). \tag{2.2}$$

The full dissipator defined in Eq. 1.2 is the expected value of the bundled dissipator:

$$\mathscr{D} \equiv \mathbb{E}\mathscr{D}_1. \tag{2.3}$$

The fluctuations in $\mathscr{D}_1$ can be mitigated by statistical sampling, using $M > 1$ random vectors $\boldsymbol{r}_m$, $m = 1, \ldots, M$, generating the bundled operators, $\mathcal{R}_1, \ldots \mathcal{R}_M$

$$\mathcal{R}_m = \sum_{\omega \in B} \frac{r_m^{\omega}}{\sqrt{M}} \mathcal{L}_{\omega}, \tag{2.4}$$

from which we define a stochastically bundled dissipator of size $M$:

$$\mathscr{D}_{1\ldots M} \rho = \frac{1}{2} \sum_{m=1}^{M} \left(\left[\mathcal{R}_m \rho, \mathcal{R}_m^{\dagger}\right] + \left[\mathcal{R}_m, \rho \mathcal{R}_m^{\dagger}\right]\right). \tag{2.5}$$

The full dissipator is still the expected value of the random dissipator: $\mathscr{D} \equiv \mathbb{E}\mathscr{D}_{1\ldots M}$ but $\mathscr{D}_{1\ldots M}$ has a smaller fluctuation, by a factor $M^{-1/2}$, than $\mathscr{D}_1$.

Once we have sampled the dissipator $\mathscr{D}_{1\ldots M}$ we use it in the following stochastic LME

$$\dot{\rho}_{1\ldots M}(t) = \frac{1}{i\hbar}\left[\mathcal{H}, \rho_{1\ldots M}(t)\right] + \mathscr{D}_{1\ldots M} \rho_{1\ldots M}(t), \tag{2.6}$$

to evolve the density matrix from its known initial value to all time values. The evolving $\rho_{1\ldots M}(t)$ is now a random density matrix with which we can estimate the exact deterministic $\rho(t)$ obtained from the LME with the full dissipator (Eq. 1.1).

2.2. **Bias mitigation via Jackknife resampling.** The stochastic fluctuations in $\mathscr{D}_{1\ldots M}$, which are proportional to $M^{-1/2}$ will cause the expected values of $\rho_{1\ldots M}(t)$ to differ from $\rho(t)$, resulting in a "bias"-ed estimator:

$$\Delta\rho_{1\ldots M}^{direct}(t) \equiv \mathbb{E}\rho_{1\ldots M}(t) - \rho(t), \tag{2.7}$$

where the superscript *direct* emphasizes the direct use of $\rho_{1\ldots M}(t)$ in the estimation of $\rho(t)$. The bias occurs because the noise in the bundled dissipator enters the evolving density matrix in a nonlinear way.

The bias typically drops as $M^{-1}$ for large sample sizes $M$ [61] and we can reduce the bias further by combining results from subsets of the bundled operators. This suite of techniques are *jackknife estimators* and they hold potential to reduce the bias faster than $M^{-1}$ [62]. For example, taking an even sample size $M$, the "jackknife$_1$" estimator is

$$\rho_{1\ldots M}^{\text{jackknife}_1}(t) = 2\rho_{1\ldots M}(t) - \rho_{1\ldots M/2}(t) \tag{2.8}$$

where $\rho_{1\ldots M}$ was defined above and $\rho_{1\ldots M/2}$ is the estimator obtained from the stochastic LME using the dissipator $\mathscr{D}_{1\ldots M/2}$, based on $\mathcal{R}_1, \ldots, \mathcal{R}_{M/2}$. The required numerical work in applying jackknife$_1$ is $\frac{3}{2}$ times that of the direct method. This is because we need to solve the LME twice, once with the $\mathscr{D}_{1\ldots M}$ dissipator and once with the $\mathscr{D}_{1\ldots M/2}$ dissipator and then apply Eq. 2.8. A more elaborate yet balanced approach is the "jackknife$_2$" estimator

$$\rho_{1\ldots M}^{\text{jackknife}_2} = 2\rho_{1\ldots M} - \frac{1}{2}\left(\rho_{1\ldots M/2} + \rho_{M/2+1\ldots M}\right). \tag{2.9}$$

The required numerical work in applying jackknife$_2$ is 2 times that of the direct method. This is because we need to solve the LME twice, once with the $\mathscr{D}_{1\ldots M}$ dissipator and once with the $\mathscr{D}_{1\ldots M/2}$ dissipator and then apply Eq. 2.8.

## 3. Results

3.1. **Anharmonic oscilator-spin system.** In order to benchmark the stochastic bundling method, we construct a hierarchy of systems where the number of Lindblad operators increases significantly, but the underlying dynamics remain largely unaffected. This approach allows us to readily identify any divergence of the bundling method with growing Hilbert space size, as other system properties are kept relatively constant. Specifically, we consider an anharmonic oscillator coupled to a qudit: spin $s = 0$ (spinless), $s = \frac{1}{2}$ (qubit), $s = 1$ (qutrit or photon), and $s = \frac{3}{2}$ (ququart).

3.1.1. *The Hamiltonian.* The Hamiltonian is a sum of tensor product operators, comprising $(2s+1) \times (2s+1)$ matrices (see Methods section) and oscillator Hilbert space operators:

$$\begin{aligned} \mathcal{H}_{sys} &= \sigma_0^{(s)} \otimes \mathcal{H}_0 + \frac{1}{2}\Gamma\sigma_z^{(s)} \otimes \mathcal{I} + \alpha\sigma_x^{(s)} \otimes \mathcal{X}_0 \\ \mathcal{H}_0 &= \mathcal{K} + \mathcal{U}, \end{aligned} \tag{3.1}$$

where $\Gamma$ is the spin gap parameter, $\alpha$ is the spin-oscillator coupling constant, and the oscillator Hamiltonian $\mathcal{H}_0$ includes kinetic energy $\mathcal{K}$ and potential energy $\mathcal{U}(\mathcal{X}_0)$. Here, $\mathcal{X}_0$ is the oscillator position operator, and the potential is defined as:

$$\mathcal{U}(x) = \max\left[U_{max}, V_\infty\left(1 - e^{-ax}\right)^2\right]. \tag{3.2}$$

Further details regarding the model are provided in the Methods section.

We assume that the system operator coupling to the environment is the oscillator position $\mathcal{X} = \sigma_0^s \otimes \mathcal{X}_0$. Neglecting the Lamb shift, we model the spectral density $\gamma(\omega)$ as:

$$\gamma(\omega) = \gamma_* e^{-\frac{1}{2}\left(\frac{\omega}{\omega_c}\right)^2} e^{\frac{1}{2}\frac{\hbar\omega}{k_B T}} \tag{3.3}$$

with $\omega_c$ being a high-frequency cutoff ($k_B$ and $\hbar$ are Boltzmann's and Planck's constants, respectively; see Supplementary Material for values of other parameters).

The Hamiltonian corresponding to spin $s$ are given in Eq. 3.1 where the $(2s+1) \times (2s+1)$ matrices are:

Table 1. Parameters of the calculation (we use atomic units).

| | | |
|---|---|---|
| $m$ | 1 | Oscillator mass |
| $[x_0, x_f]$ | $[-10, 20]$ | Real space grid parameters Eq. 3.4 |
| $N_x$ | 30 | |
| $\Delta x$ | 1 | |
| $V_\infty$ | 4 | Morse potential parameters, Eq. 3.2 |
| $U_{\max}$ | 6 | |
| $a$ | 0.2 | |
| $\omega_c$ | $\sqrt{2}$ | Coupling function parameters, Eq. 3.3 |
| $\gamma_*$ | 0.005/0.02 | |
| $k_B T$ | 0.25/1 | |
| $\xi$ | 3.4/0.7 | Initial state, Eq. 3.7 |
| $\delta t$ | 0.25/0.125 | Runge-Kutta time step. The larger $\gamma_*$ the smaller $\delta t$ |
| $\Gamma$ | 0.1 | spin gap and oscillator-spin coupling Eq. 3.1 |
| $\alpha$ | 0.1225 | |

- $\sigma_0^{(s)} = \mathbb{1}_{2s+1}$ is the $(2s+1)$-dimensional unit matrix
- $\sigma_z^{(s)}$ is the diagonal matrix with entries $-1, -\left(\frac{s-1}{s}\right), \ldots, \frac{s-1}{s}, 1$ on the main diagonal
- $\sigma_x^{(s)}$ is a tridiagonal matrix with 0's on the main diagonal and 1's on the sub- and super-diagonals.

The oscillator itself has a Hilbert space consisting of continuous wave functions $\psi(x)$ supported on the interval $x \in [x_0, x_f]$ (vanishing outside this interval). We discretize these functions, representing them as $N_x$-dimensional vectors $\psi_n \equiv \psi(x_n)$ where

$$x_n = x_0 + n\Delta x, \quad n = 0, \ldots, N_x, \tag{3.4}$$

are grid points and $N_x$ is the grid size (see Table 1 for values of the various parameters).

The position operator $\mathcal{X}$ and potential $\mathcal{U} \equiv U(\mathcal{X})$ act as follows: $(\mathcal{X}\psi)_n = x_n \psi_n$ and $(\mathcal{U}\psi)_n = U(x_n)\psi_n$ respectively, where $U(x)$ is the Morse potential given in Eq. 3.2.

The kinetic energy operator $(\mathcal{K}\psi)_n = -\frac{\hbar^2}{2m}(\psi'')_n$, is approximated using a sixth-order (seven-point) finite-difference approximation for the second derivative:

$$(\psi'')_n = \frac{1}{\Delta x^2} \sum_{m=-3}^{3} a_{|m|} \psi_{n+m}, \tag{3.5}$$

where

$$a = \left(-\frac{49}{16}, \frac{3}{2}, -\frac{3}{20}, \frac{1}{90}\right) \tag{3.6}$$

, combined with the boundary condition $\psi_{-k} \equiv \psi_{N_x+k} \equiv 0$ for all positive $k$.

The potential and lowest-lying energy levels of the oscillator's Hamiltonian are shown in the left panel of Figure 3.1.

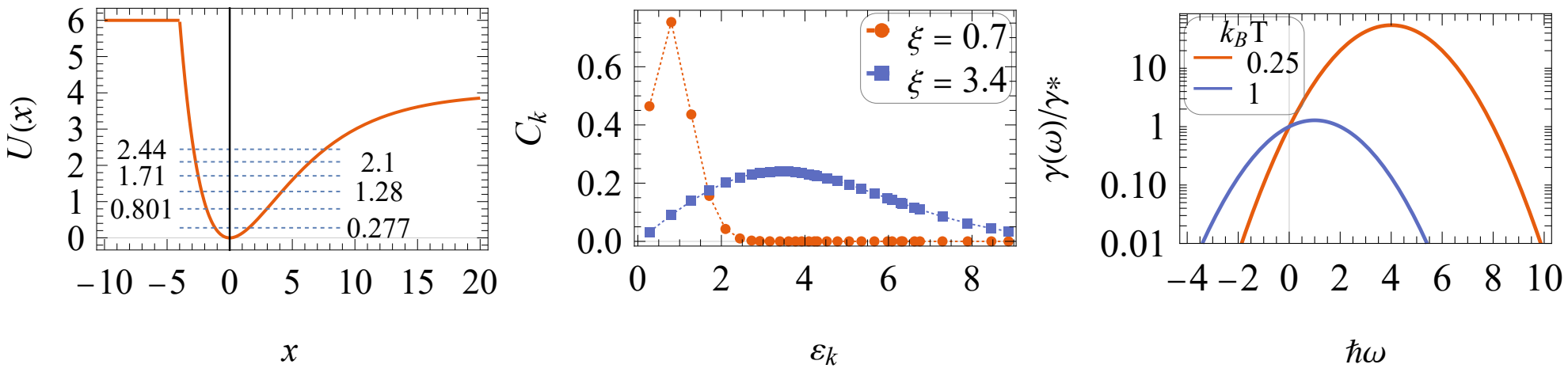

FIGURE 3.1. Left: The Morse potential $U(x)$ used in this example. The dashed lines indicate the six low-lying eigenvalues (the total number of eigenvalues is 31). Middle: the coefficients $C_k$ for the $k^{th}$ eigenstate of energy $\varepsilon_k$ of the cold and hot initial states of the spinless system (see Eq. 3.7). Right: The coupling function $\gamma(\omega)$ (Eq. 3.3) for the two temperatures used in this study.

3.1.2. *The initial states.* We take the initial density matrix as a pure state $\rho_0 = \frac{|\psi_0\rangle\langle\psi_0|}{\langle\psi_0|\psi_0\rangle}$, where

$$|\psi_0\rangle = \sum_{k=0}^{N_x} |k\rangle C_k, \qquad , C_k = \varepsilon_k e^{-\frac{\varepsilon_k^2}{2\xi^2}} \tag{3.7}$$

where $|k\rangle$ and $\varepsilon_k$ are the eigenstates and eigenvalues of the total Hamiltonian and the parameter $\xi$ takes two values: 0.7 ("cold"), and 3.4 ("hot"). The coefficients $C_k$ corresponding to these values are shown in the middle panel of Figure 3.1.

3.1.3. *The Lindblad equation propagator.* The fourth-order Runge-Kutta propagator with a time step of $\delta t$ (see Table 1) is used to solve the Lindblad Master Equations starting from the initial pure state, obtaining

$$\rho_n = \rho(n \times \Delta t) \tag{3.8}$$

where $\Delta t = 1\hbar/E_h$ and $n = 1, 2, \ldots$. We record the energy $\mathcal{E}_n = \mathrm{Tr}[\rho_n \mathcal{H}_\ell]$, position $\mathcal{X}_n = \mathrm{Tr}\left[\rho_n\left(\sigma_0^\ell \otimes \mathcal{X}\right)\right]$ and purity $\mathcal{P}_n = \mathrm{Tr}\left[\rho_n^2\right]$ transients. The parameter $\delta t$ is small enough to bring all solutions to convergence of 9 significant digits.

3.2. **The cooling and heating dynamics.** Figure 3.2 illustrates the dynamics of our system under two distinct scenarios and coupling regimes:

**Cooling Scenario** (Figure 3.2, left panels): A hot, spinless oscillator ($\xi = 3.4$) is coupled to a low-temperature bath ($k_BT = 0.25$) with environmental coupling parameters $\gamma_* = 0.02$ and $\gamma = 0.005$. The energy ($\mathcal{E}$) decays monotonically towards its equilibrium value, exhibiting faster relaxation in the strong coupling regime. The oscillator position ($\mathcal{X}$) initially expands rapidly from 4 to approximately 8 (strong coupling) or 9 (weak coupling) before slowly contracting to its equilibrium value of $\mathcal{X}_{eq} \approx 0.36$. The purity experiences a rapid initial drop from unity, followed by a prolonged period at $\mathcal{P} \approx 0.05$, finally increasing towards the thermal equilibrium value of $\mathcal{P} \approx 0.76$ as the system approaches its ground state.

**Heating Scenario** (Figure 3.2, right panels): A cool ($\xi = 0.7$) oscillator coupled to a spin-$\frac{1}{2}$ is immersed in a hot bath ($k_BT = 1$). The energy increases monotonically, albeit with slower timescales compared to the cooling scenario. In the weak coupling regime, the position exhibits coherent oscillations at intermediate times ($t \approx 100$) as it approaches its thermal equilibrium length. The purity decays monotonically towards its equilibrium value of $\mathcal{P} \approx 0.2$.

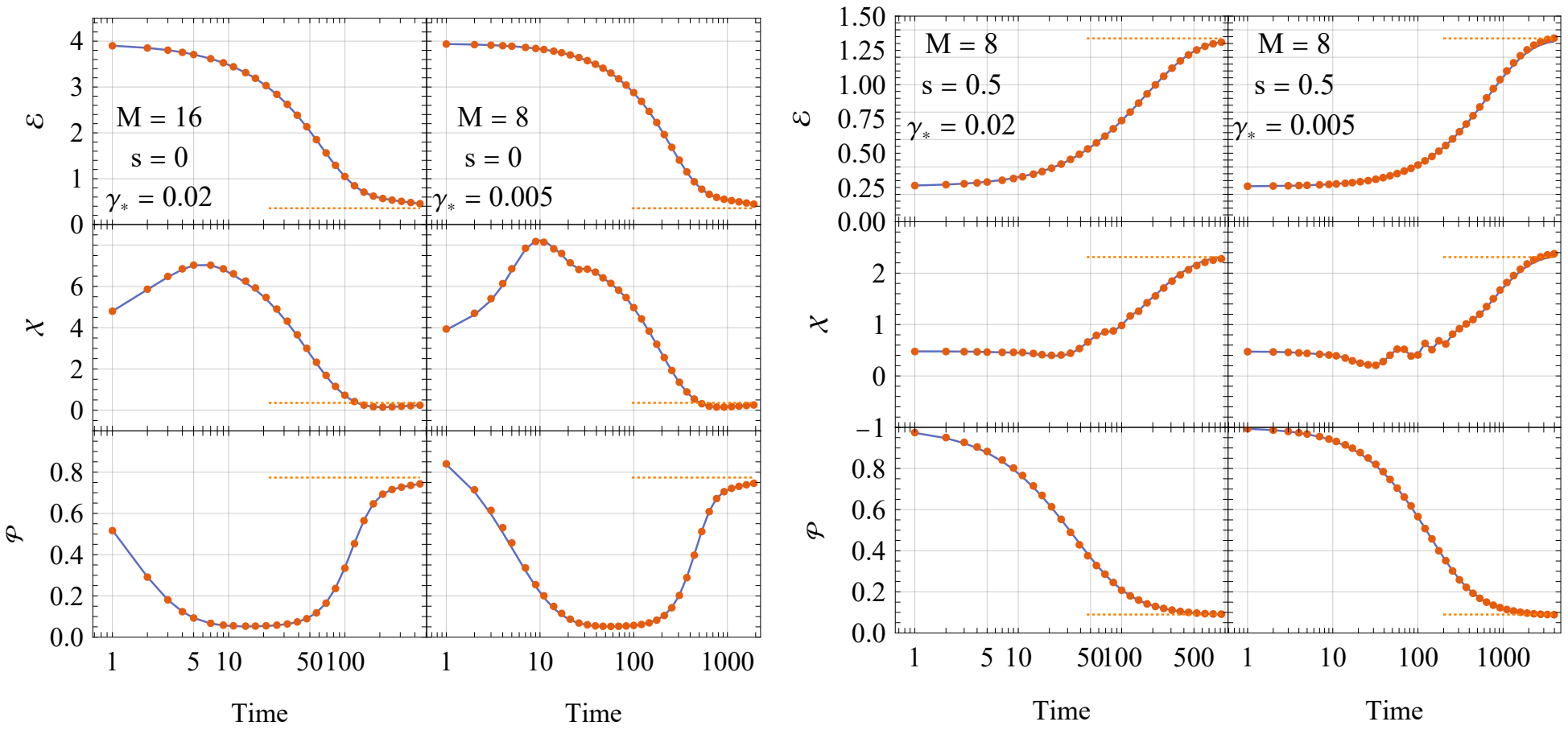

FIGURE 3.2. Energy ($\mathcal{E}$), position ($\mathcal{X}$), and purity ($\mathcal{P} = \mathrm{Tr}\rho^2$) transients for a Morse oscillator initially in a pure state (Eq. 3.7). Left panels: oscillator without spin ($s = 0$) undergoing cooling ($\xi = 3.4$, $k_BT = 0.25$). Right panels: oscillator coupled to a spin $s = \frac{1}{2}$ experiencing heating ($\xi = 0.7$, $k_BT = 1$). Deterministic (solid lines) and stochastic (markers) dynamics are compared. Dashed lines indicate thermal equilibrium values. The deterministic dynamics are based on 753 (left panels) and 3287 (right panels) Lindblad operators. The stochastic dynamics in both panels use 8 bundled operators.

3.3. **The stochastic calculation.** The stochastic approach estimates the expectation value of an observable $\langle \mathcal{O} \rangle_t$ at time $t$ as $\langle \mathcal{O}_{1\ldots M} \rangle_t = \mathrm{Tr}\left[\rho_{1\ldots M}(t)\mathcal{O}\right]$, where $\rho_{1\ldots M}(t)$ is the solution at time $t$ of the LME with a stochastic dissipator (Eq. 2.6) having $M$ bundled Lindblad operators. Figure 3.2 shows the expectation value estimates of the energy, position, and purity based on an $M = 8$ stochastic calculation. These estimates closely approximate those of the full deterministic calculation, which utilizes a dissipator with over 725 (spinless oscillator) and 3287 (oscillator coupled to spin $\frac{1}{2}$) Lindblad operators, despite being approximately two orders of magnitude faster to compute for the spinless case, and nearly 400 times faster for the case with spin.

The computational speedup achieved through the stochastic approach comes at the cost of reduced accuracy. To quantify this, we examine the maximum root-mean-square error (RMSE) for energy, $\Delta\mathcal{E}$, and position, $\Delta\mathcal{X}$, in the cooling scenario with coupling strength $\gamma_* = 0.02$, as depicted in Figure 3.3. The definition of the RMSE and the methods used for its determination are detailed in the Methods section below. Results for the other scenarios exhibit similar trends and are presented in the Supplementary Material.

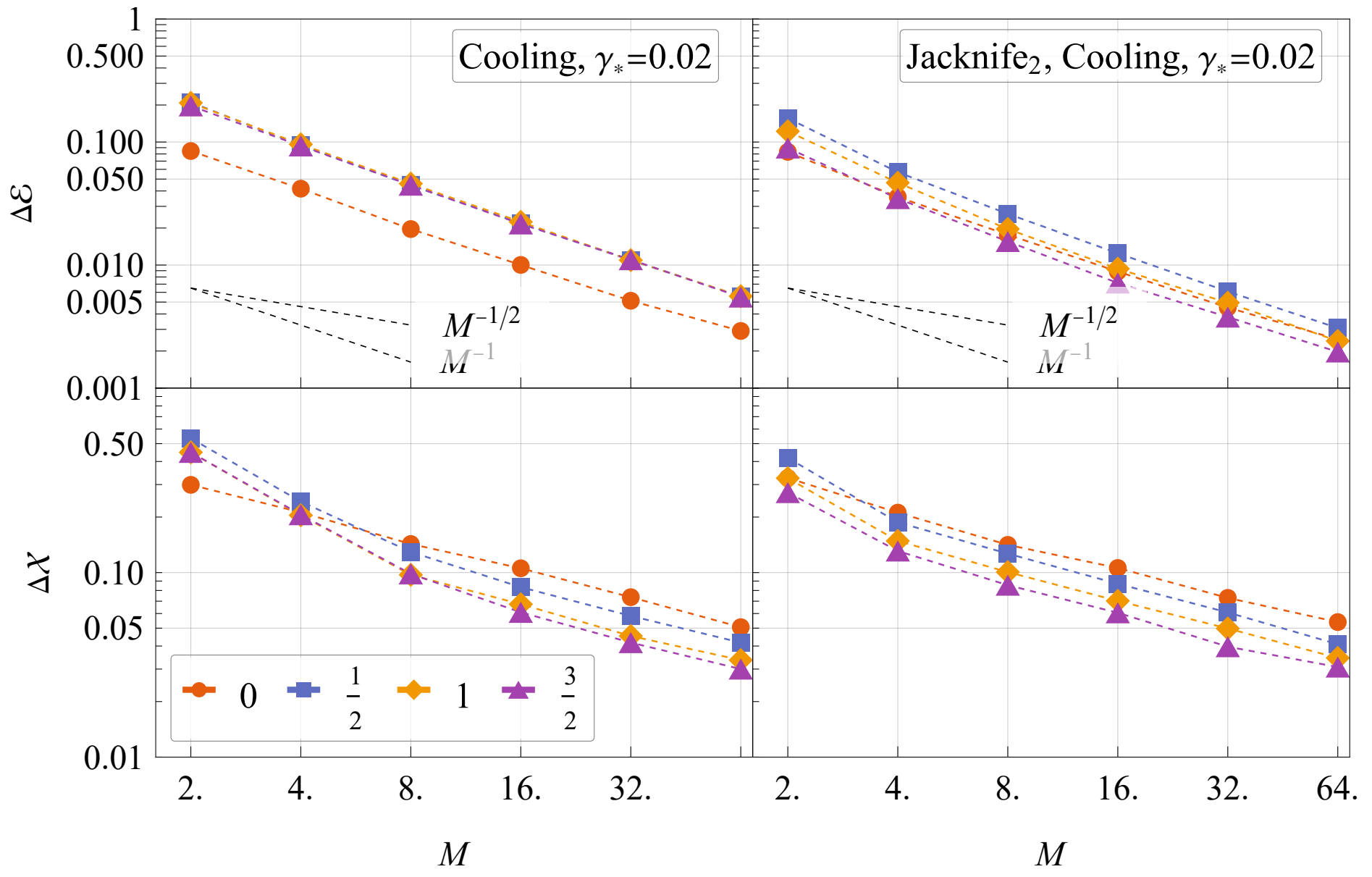

FIGURE 3.3. Maximal RMSEs in energy and position for open spin-oscillators coupled to a cold environment ($k_B T = 0.25$) shown as functions of the number of stochastically bundled Lindblad operators, M. Calculations start from an initially pure and hot state. Dashed lines indicate M$^{-1}$ and M$^{-1}/^{2}$ slopes.

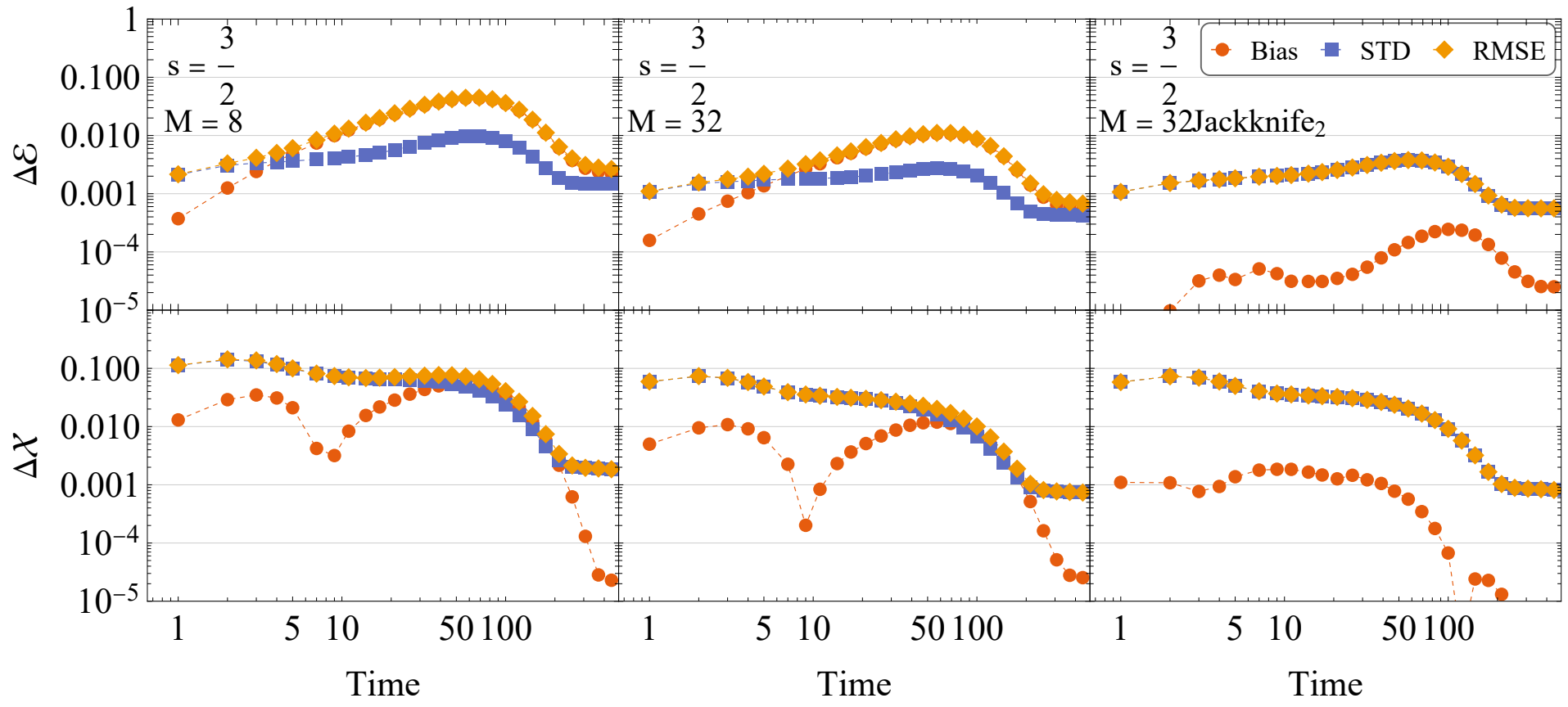

FIGURE 3.4. The errors in energy (top) and position (bottom) for the cooling scenario of a spin-3/2 oscillator system with $\gamma_* =$ 0.02. The left panels show errors for a bundled dissipator with 32 Lindblad operators, the middle panels for 8 operators, and the right panels for 8 operators within the jackknife approach of Eq. 2.9.

For energy, the lowest $\Delta\mathcal{E}$ values occur for the spinless system. Increasing the spin to $s = \frac{1}{2}$, where the Hilbert space dimension increases by a factor of 2 and the number of Bohr frequencies by a factor of 4, $\Delta\mathcal{E}$ grows by a factor of 2.5. Further increases to $s = 1$ and $s = \frac{3}{2}$, doubling once more the Hilbert space dimension, do not significantly change $\Delta\mathcal{E}$ further. The maximal RMSE in position, $\Delta\mathcal{X}$, exhibits different behavior: $\Delta\mathcal{X}$ decreases with increasing spin and Hilbert space dimension. We conclude that consistently increasing Hilbert space dimensions does not significantly affect the statistical errors. Therefore, the number $M$ of stochastically bundled Lindblad operators can remain constant as the Hilbert space dimension $N$ grows, without sacrificing accuracy.

Next, we discuss the dependence of the maximal RMSEs on $M$. While $\Delta\mathcal{E}$ decreases proportionally to $M^{-1}$, $\Delta\mathcal{X}$ behaves differently. For non-zero spin and $M \leq 8$, $\Delta\mathcal{X}$ decreases as $M^{-1}$, and in other cases ($M > 8$) it decreases as $M^{-1/2}$.

Analysis of the Mean Squared Error (MSE) provides insight into the observed behavior. The MSE is decomposable into the sum of the variance and the squared bias, with definitions and estimation procedures for these components detailed in the Methods section.

Statistical theory establishes that both bias and variance exhibit scaling relationships with sample size ($M$), scaling as $M^{-1}$ [63]. Consequently, when the MSE is dominated by variance, its root, the Root Mean Squared Error (RMSE), will be approximately equal to the standard deviation, thus scaling proportionally to $M^{-1/2}$. Conversely, when the MSE is dominated by the squared bias, the RMSE approximates the bias, scaling as $M^{-1}$.

Therefore, the dominant error source within the MSE can be inferred from the RMSE scaling behavior: $M^{-1}$ scaling indicates bias dominance, whereas $M^{-1/2}$ scaling signifies variance dominance. This relationship is further corroborated by the effect of the jackknife resampling method on the maximal RMSE, as illustrated in Figure 3.3. When bias dominates, the jackknife effectively reduces the maximal RMSE. In contrast, when variance dominates, the jackknife exerts a negligible effect on the maximal RMSE.

To demonstrate this effect further, Figure 3.4 presents the time-dependent RMSE, fluctuations (standard deviation), and bias for energy and position observables at $s = \frac{3}{2}$, with results for $M = 8$ (left) and $M = 32$ (right). For the energy observable, the bias error generally dominates the RMSE, particularly at the maximal RMSE occurring around $t = 50$. This observation confirms the $M^{-1}$ scaling of the maximal RMSE for energy. In contrast, the maximal RMSE for the position observable ($\Delta\mathcal{X}$), observed at early times, is dominated by fluctuation error, consistent with the expected $M^{-1/2}$ scaling.

As previously discussed, solving the Lindblad equation has a formal algorithmic complexity of $\mathcal{O}(N^5)$ per time step, where $N$ is the Hilbert space dimension. Our results demonstrate that the number $M$ of bundled Lindblad operators is largely independent of $N$. This independence implies a significant reduction in complexity to $\mathcal{O}(N^3)$. This improved scaling is confirmed by our calculations, as shown in Figure 3.5. The $\mathcal{O}(N^5)$ scaling of deterministic calculations quickly becomes computationally prohibitive. In contrast, our stochastic approach can handle systems described by hundreds of thousands of Lindblad operators in the traditional formalism, effectively reducing the computational cost.

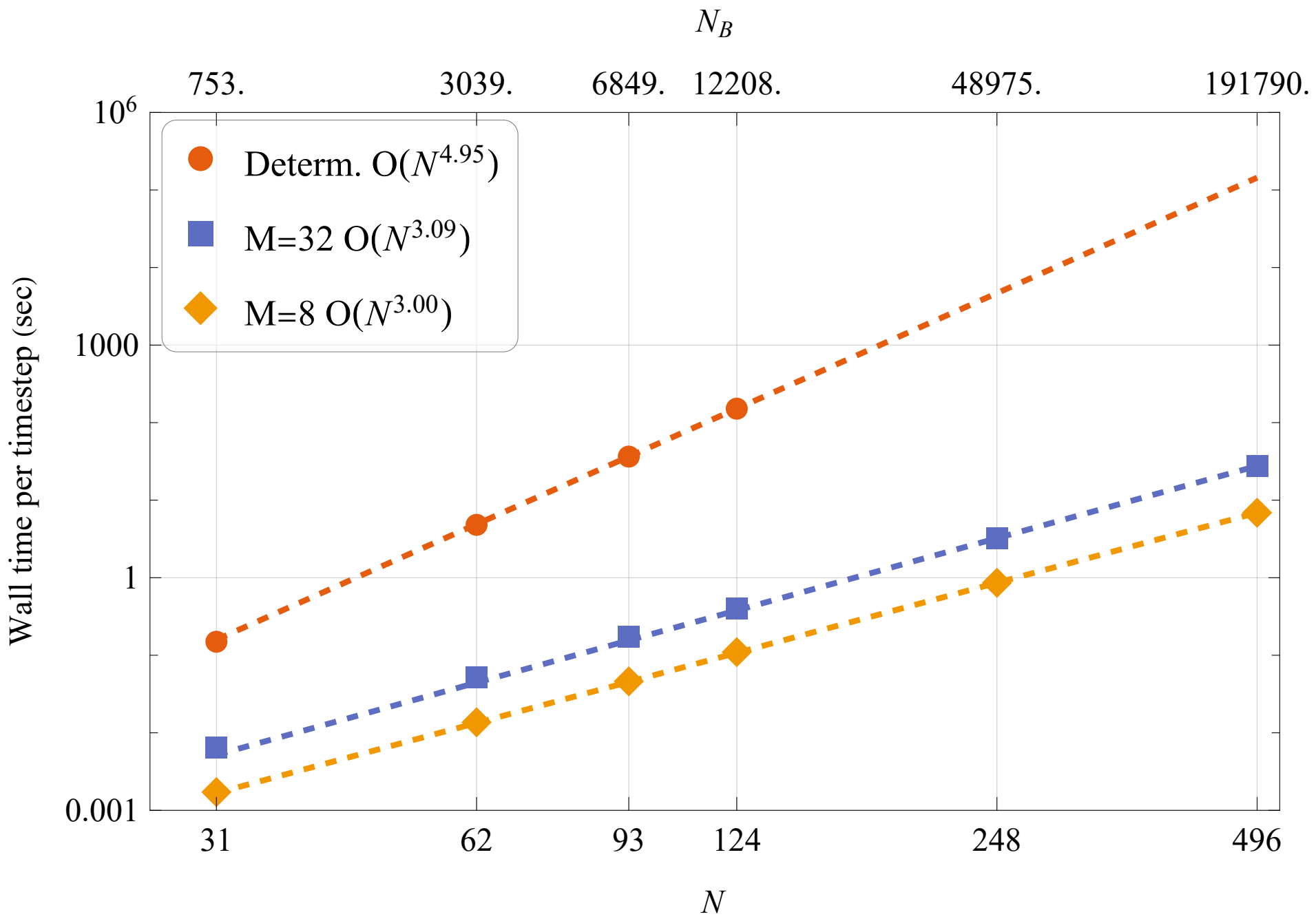

FIGURE 3.5. Wall times for oscillator-spin systems as function of the Hilbert space dimension $N$ (and the corresponding number of Bohr frequencies, $N_B$) . The scaling of the calculations, $n$, is determined by fitting the data to $t = aN^n$.

## 4. Discussion

The stochastic bundling method, introduced in this paper, offers a powerful new approach to calculating the dissipator in the Lindblad master equation, a key challenge in the study of open quantum systems. By representing the dissipator with a small number of stochastically bundled Lindblad operators, this method effectively reduces the complexity of the calculation.

Our numerical experiments, including the example of a Morse oscillator coupled to a spin-1/2 particle, demonstrate the effectiveness of this method. We were able to closely approximate the system's dynamics using only $M = 8$ bundled operators instead of the full set of 3069 Lindblad operators, resulting in a 400-fold reduction in computation time. This highlights the potential of stochastic bundling to address the computational bottleneck associated with large open quantum systems.

While significantly reducing computational cost, it is crucial to assess the accuracy of the stochastic bundling method. The accuracy is influenced by factors such as the specific system under investigation, the desired accuracy, the number of bundled operators employed, and the application of jackknife resampling. Notably, our findings demonstrate that the number $M$ of bundled operators required to attain a given accuracy level does not scale with the Hilbert space dimension, $N$. This leads to a significant reduction in computational scaling from $\mathcal{O}(N^5)$ for traditional deterministic methods to effectively $\mathcal{O}(N^3)$ for our method, as $M$ is typically small and independent of $N$. This scaling advantage makes the stochastic bundling method particularly well-suited for large systems with high-dimensional Hilbert spaces.

We have also demonstrated the effectiveness of jackknife resampling in mitigating bias. This is crucial for interpreting our method as a physically meaningful approximation. When the bias, introduced by using a finite number of bundled operators, is sufficiently small, the random density matrix $\rho_{1...M}(t)$ (obtained from Eq. 2.6) can be interpreted as representing a plausible trajectory of the system's mixed state under environmental fluctuations. While fluctuation error remains it can be mitigated through repeated calculations and averaging, as the fluctuations tend to cancel out.

This approach bears resemblance to unraveling via quantum state diffusion (QSD) processes [45, 64], which involve solving a stochastic Schrödinger equation (SSE) that yields evolving pure states. Each of these pure states represents a plausible evolution of the system under the influence of a fluctuating environment, and averaging over them produces the exact density matrix $\rho(t)$. Crucially, our unraveling differs from QSD in that our random states are inherently mixed. Furthermore, these mixed states exhibit significantly smaller fluctuations compared to the pure states in QSD. This difference suggests that our method may offer advantages in terms of stability and efficiency for certain systems, potentially due to the reduced sensitivity to numerical errors arising from the smaller fluctuations.

While the stochastic bundling method offers significant advantages for studying open quantum systems, it also has limitations that suggest avenues for future research. Currently, the method assumes a weak coupling between the system and its environment. As the coupling strength increases, a larger number of bundled operators, M, may be required to maintain accuracy. Investigating the method's performance in the strong-coupling regime is therefore crucial. One potential approach is to employ the reaction coordinate method [59, 60], which effectively maps a strongly coupled system to a larger, weakly coupled one. Another limitation is the method's restriction to Markovian systems, which lack memory effects. Extending the stochastic bundling technique to non-Markovian systems presents a significant challenge, primarily because the dissipator's definition becomes considerably more complex, and the applicability of bundling is unclear. Future work should explore the possibility of extending this approach to non-Markovian systems, potentially including those described by Redfield equations.

Despite these limitations, the stochastic bundling method presents a valuable new tool for understanding open quantum systems, especially those with high dimensionality. This dramatic improvement in computational efficiency, stemming from the reduced scaling, enables the simulation of much larger and more complex quantum systems than previously feasible. The ability to explore these systems opens up new avenues for investigating a wide range of phenomena in quantum optics, condensed matter physics, and quantum information science. The stochastic bundling method, therefore, not only addresses a critical computational bottleneck but also expands the scope of quantum systems amenable to theoretical investigation.

**Acknowledgement.** We gratefully acknowledge the support of the Israel Science Foundation under grant ISF-1153/23.